\documentstyle[12pt,aasms4]{article}

\begin{document}

\title{Ultraviolet structure in the lensed QSOs 0957+561
\footnote{Based on observations with the NASA/ESA \it Hubble Space 
Telescope, \rm obtained at the Space Telescope Science Institute, which 
is operated by AURA Inc under NASA contract NAS5-26555 }
\footnote{Guest User, Canadian Astronomy Data Centre, which is operated
by the Dominion Astrophysical Observatory for the National Research Council
of Canada's Herzberg Institute of Astrophysics}} 

\author{J.B. Hutchings}
\affil{Herzberg Institute of Astrophysics, NRC of Canada,\\ Victoria, B.C.
V9E 2E7, Canada; john.hutchings@nrc.ca}

\begin{abstract}

   Imaging and spectra of the lensed QSO pair 0957+561 are presented and
discussed. The data are principally those from the STIS NUV MAMA, and
cover rest wavelengths from 850\AA~ to 1350\AA. The QSOs are both extended
over about 1 arcsec, with morphology that matches with a small rotation,
and includes one feature aligned with the VLBI radio jets. This is the first
evidence of lensed structure in the host galaxy. The off-nuclear 
spectra arise from emission line gas and a young stellar population. The gas 
has velocity components with radial velocities at least 1000 km s$^{-1}$ 
with respect to the QSO BLR,
and may be related to the damped Ly$\alpha$ absorber in the nuclear spectra.

\end{abstract} 


\section{Introduction and observations}

   Q0957+561 is a bright gravitationally lensed QSO pair at redshift 1.415.
The two images are 6.2" apart, and deep imaging reveals a chief lensing
galaxy between them, which lies in a cluster at redshift 0.36. The
lensing model and QSO host galaxy morphology have been discussed recently
by Keeton et al (2000), and their paper contains a good bibliography of
related earlier work. The QSO UV spectra reveal that they have damped
Ly$\alpha$ absorption and other associated absorbers that arise in the
QSO host galaxy (see Michalitsianos et al 1997, and Dolan et al 2000).
Repeated UV nuclear spectroscopy has revealed changes and differences in the
UV absorbers that are discussed in the above references.

   In order to look for extended Ly$\alpha$ in the vicinity of the QSO, and
to study the structure of the nuclear regions, direct images and slitless
spectra were taken with the NUV detector of the Space Telescope
Imaging Spectrograph (STIS). Both QSO images were in the field of view. 
Table 1 summarizes the observations, and also
others taken with the same STIS grating, which were obtained from the
HST archive. This paper discusses the non-nuclear spectra and UV imaging,
which have not been covered in the other work referred to above.

\section{UV images}

    The UV images are shown in Figure 1, along with images of stars taken
with the same detector. For comparison, several star images, with 
exposure times and signal levels similar to the QSO images, were taken 
from the HST archive. While the star images all show deviations from 
circular symmetry, due to HST tracking jitter, they are all smaller and
rounder than the QSO images, as can be seen in Figure 1 and also from
the luminosity profiles in Figure 2. Figure 3 shows contours of the two
QSO images.

   Figure 1 shows both QSOs, rotated to the conventional orientation, the 
most non-circular star image, and the average image from several stars. 
The images have been smoothed with a gaussian of $\sigma$=2 `pixels' (0.025")
to reduce the granularity of the photon-counting MAMA detector, but this
does not significantly affect the spatial resolution (FWHM about 5 pixels) 
which is determined by the HST optics.
The QSOs both have very similar `kidney-shaped' inner profiles, and fainter
radial structures, and match each other best with a small relative 
rotation, depending on whether the cores or radial spokes are
fit. No such structure is seen in any star image, and the jitter in the
QSO observation does not indicate a difference from the star observations.
Thus, we conclude that there is resolved structure in the inner 0.3 arcsec
diameter of the QSO images, and that the lensing distortion and rotation within
this spatial scale is small. The radial structures are seen out to distances 
of about 0.4 arcsec, and do show larger differences in extent and relative
brightness. 

   The luminosity profiles are derived from the IRAF `ellipse' task, and
plot the semi-major axis of the best fit ellipse to each signal level. The
star profile is the mean profile from all the star images. The two spatial
axis plots show that these inner profiles are not exponential or power-law.
The QSO profiles are resolved at essentially all radii: the kink in the star
profiles is caused by the HST PSF structure that shows up most clearly in
the unresolved stellar images. We can also see that the B image is larger
than the A image in the inner parts (by some 20\% in radius, normalised
to the peaks). The B image is brighter, but only by 1\%, as the peak
signal is lower. We discuss the image morphology in the last section.

   There is no sign of the lensing galaxy or any other object in the
25 arcsec field of the UV image, as expected for the redshifted SED
of a cluster galaxy at redshift 0.36. The only WFPC2 images of the QSOs have
0.1" pixels and show no resolved structure, but do reveal the lensing galaxy.

\section{Slitless spectra}

    The dispersed image contains spatially resolved spectra of both
QSO images. As these have no wavelength calibration, it was necessary
to process the raw image into one with spatial and wavelength axes, separately
for each QSO. The wavelength scale was established by using the nuclear
emission lines and assuming the QSO emission-line redshift of 1.415. These 
are described together with the standard long-slit spectra (see Table 1)
in the next section. Figure 4 shows the nuclear spectra, and off-nuclear
spectra, also described in the next section. As the nuclear UV spectra have 
been discussed in detail by the papers noted above, this paper does not 
investigate them further. 

   The dispersed image was inspected carefully for signs of Ly$\alpha$
emission elsewhere in the field. This might arise from emission-line
gas at the QSO redshift (i.e. associated with the host galaxy), either
on the scale of the host galaxy, or close to the QSO nucleus. The lensing
of the QSO and environments means that this may be found anywhere
between the two QSO nuclei and at some distance beyond. The
background is remarkably uniform, and no significant extra flux is
seen in the likely regions, compared with the rest of the image. Close
to the nuclei, Ly$\alpha$ emission is not noticeably
extended beyond the rest of the UV continuum. Thus, the data do not reveal
any detached Ly$\alpha$ emission line gas around the QSO.

\section{QSO off-nuclear spectra}

   The cross-sections of the UV spectra are shown in Figure 5. The resolved
width of the nuclear regions cause the slitless spectra to be wider than
the  0.2" slit spectra. Further, the extended structure seen in the UV
images in Figure 1, will give rise to more off-nuclear signal in the slitless
spectra. Both these are seen. The closest comparison is between the
two spectra of QSO A with dispersion angles 110$^o$ and 114$^o$. There are
additional uncertainties in the exact positioning of the nuclear region
with the 0.2" slits.

   It was found that the regions outside 0.1" in the slit spectra and outside
0.21" in the slitless spectra are free of nuclear contamination, by noting that
they do not show the damped Ly$\alpha$ absorption. The regions inside that in 
the slitless spectra (dotted lines in Figure 4), are slightly contaminated,
and Figure 3 shows the effect of subtracting a scaled nuclear spectrum just 
sufficient to remove the damped Ly$\alpha$ absorption.

   The off-nuclear spectra for components A and B in the slitless data
are very similar, extracted as shown in Figure 5. Thus, Figure 3 shows the
combined A+B spectra. There is clearly off-nuclear Ly$\alpha$, which
we discuss in more detail below, but we also note the possible presence
of N V (1238-42\AA), and an associated UV continuum, which has the same
spatial extent as the line emission. Both suggest the presence of a young
stellar population. Extraction of off-nuclear spectra from the G430L and G750L
observations (see Table 1), shows a noisy broad emission at redshifted
C IV (1550\AA) which also suggests a distributed young population. At longer
wavelengths, the off-nuclear spectra have a flater (i.e. redder) SED, which
presumably
arises from an older stellar population as well. At the QSO redshift, we can
see rest wavelengths out to about 4000\AA, but the signal is too weak and
noisy to reveal line features.

   Figure 6 shows the Ly$\alpha$ emission separately for all off-nuclear
spectra, and both QSO components. The combination of the spatial extent
of the off-nuclear emission-line regions, the slit width (or lack of slit),
and the uncertain placement of the nucleus within the slit, make velocity
measures of the line rather hard to interpret.  However, the width and
offset of the lines from the QSO redshift (shown dotted in) indicates
that some Ly$\alpha$ emission gas has velocities at least of order 
1000 km s$^{-1}$ with respect to the nucleus. The W spectra at all orient
angles indicate redshifted emission that corresponds to radial velocities
of several thousand km s$^{-1}$. As noted above, the behaviour
in components A and B is similar and suggests that the lensing does not
distort the spatially resolved near-nuclear regions very much.

\section{Discussion}

   The far-UV rest-frame imaging and spectra of Q0957+561 show that the 
QSOs are resolved over scales of about 1 arcsec, and that the two lensed
components have very similar morphology out to radii of 0.3 arcsec, 
although component B is larger by about 20\%. 
There are also fainter radial features that extend
over diameters of 1.5 arcsec, that do show differences between A and B.
There is no sign of the lensing galaxy, or of more extended L$\alpha$ 
in the UV data.

   The radio structure is of interest, and known to comparable spatial
resolution. Conner, Lehar, and Burke (1992) published VLBI maps that
resolve jets from both A and B over some 100mas. Garrett et al (1994) 
show these in more detail, and there is larger structure mapped with the
VLA by Greenfield, Roberts, and Burke (1985). The 100mas jets are at position
angles 23$^o$ and 17$^o$ for A and B, respectively, although they are both 
near 14$^o$ in the inner 10mas, and the A jet curves to PA nearly 90$^o$
over 5 arcsec. The radio flux magnification (B/A) is close to 0.75 (see
e.g. Press and Rybicki 1998), while the optical values are near to unity.
This disagreement is considered to be due to microlensing, or possibly the
nearness  of B to a lensing caustic.

   The UV images reported here do have an extension with position angle 
about 25$^o$, over the innermost 100mas, which suggests that they are
associated with the radio jets. They do also have extensions with similar
size, but less sharpness, at angles -60$^o$ and -170$^o$, as seen in Figure 3.
 The best fit
between A and B over the inner 200mas radius is with a 4.5$^o$ rotation
clockwise of B, which is the opposite direction of the radio jet angle
differences. As noted, the UV flux magnification is 1.01 and the size
is 1.2 (both B/A). Thus, it appears that some of the UV structure is 
aligned with the radio jets, but the difference in magnification is similar
to that at optical wavelengths. 

    The off-nuclear spectra indicate that there are clouds of emission
line gas within the QSO host galaxy with relative velocities of the order
of 1000 km s$^{-1}$, and possibly significantly higher. This suggests
association with the damped absorber seen in the nuclear 
spectra, which indicates outflows of about 2500 km s$^{-1}$.
The off-nuclear Ly$\alpha$ flux is about 10$^{-12}$ erg s$^{-1}$, 
(10\% of the nuclear line BLR flux), and the equivalent widths similar 
to the nuclear at about 70\AA. 
The off-nuclear spectra also indicate the presence of a young hot stellar
population, as well as a redder population that begins to dominate 
above rest wavelengths 4000\AA.

   Further high resolution imaging and spectra should provide more 
information on the spatial structure, and possibly associate it with
changes seen in the nuclear absorption spectra. The inner morphology
also may provide further constraints on lensing models for the QSO.

\clearpage

\begin{deluxetable}{ccccccc}
\tablenum{1}
\tablecolumns{7}
\tablecaption{{STIS UV observations of Q0957+561}}
\tablehead{\colhead{QSO} &\colhead{Date} &\colhead{Slit} 
&\colhead{Grating} &\colhead{exp} &\colhead{Orient\tablenotemark{a}}
&\colhead{Note}\\
&\colhead{d/m/y} &&&\colhead{(sec)} &\colhead{(deg)}}

\startdata
A &15/04/99 &52x0.2 &G230L &1900 &81.9 &also G430L,G750L\\
B &02/06/00 &" &" &1900 &49.6 &also G430L,G750L\\
A &08/09/00 &" &" &1500 &$-$66.3\\
A+B &10/03/01 &open &25MAMA &200 &109.7\\
A+B &10/03/01 &open &G230L &2200 &109.7\\

\enddata
\tablenotetext{a}{Position angle of image y-axis degrees E of N}
\end{deluxetable}

\clearpage
\centerline{References}

Conner S.R., Lehar J., Burke B.F., 1992, ApJ, 387, L61

Dolan J.F., Michalitsionos A.G., Nguyen Q.T., Hill R.J., 2000, ApJ, 539, 111

Garrett M.A., Calder R.J., Porcas R.W., King L.J., Walsh D, Wilkinson P.N.,
1994, MNRAS, 270, 457

Greenfield P.E., Roberts D.H., Burke B.F., 1985, ApJ, 293, 370

Keeton, C.R., et al, 2000, ApJ, 542, 74

Michalitsianos A.G. et al, 1997, ApJ, 474, 598 

Press W.H., Rybicki G.B., 1998, ApJ, 507, 108

\clearpage

\centerline{Captions to figures}

1. UV images (rest frame 900-1300\AA) of the QSOs (upper), compared with 
stellar images taken at the same observed wavelengths and detector (lower).
Images are all 8" on a side, have similar signal levels, and the same
display lookup table. The QSOs are oriented N up and E to the right.
The left QSO image is component A and the right one is B. The left star
is the `worst' one found in the archive, and the right one is the composite
of four stars. Note the similar core structure in the QSOs, larger than the 
stars, and also the radial faint rays emerging from the QSO images.

2. Radial profiles of the QSO images and the mean of the star images, plotted
on R$^{1/4}$ and R scales. The QSO images are extended, with the B image being
larger, but with a very similar profile. Formal error bars from the ellipse
task are plotted.

3. Contour plots of the two UV images. North is up and East to the left.
The extension to the NE is aligned with the VLBI radio jets. The two
images align best with a rotation of 4.5$^o$.

4. Spatially resolved UV spectra of Q0957+561. The wavelength scale
is the rest frame, and spectra step across the nucleus as shown in the
cross sections in Figure 5. As the inner QSO structure (and the extracted
spectra) are so similar, the off-nuclear spectra are combined from both A and B 
components. The central panel shows the two nuclear spectra (scaled down by
some 20 times compared with the others). Note the small  
differences in the damped absorption lines. The two extractions nearest
the nucleus are probably contaminated by nuclear light, and are also shown
after nuclear subtraction sufficient to remove the Ly$\alpha$ absorption.

5. Cross-sections of all spectra in Table 1. The vertical dotted lines show
the extraction windows for the spectra in Figure 4. The lower, slit spectra 
are narrower due to the slit width. The off-nuclear spectra in Figure 6
are extracted using the light outside the inner dotted lines in the upper
panel. The different dispersion orientations and uncertainties in the
exact slit placement cause the different amounts of off-nuclear light.

6. Panel of Ly$\alpha$ emission separately for the two QSO images. These
spectra are normalised to a continuum fitted to the entire UV spectrum in
all cases. The
upper four (dashed) spectra are from the slitless data. Note the Ly$\alpha$
profile changes, and the possible appearance of N V away from the nucleus.

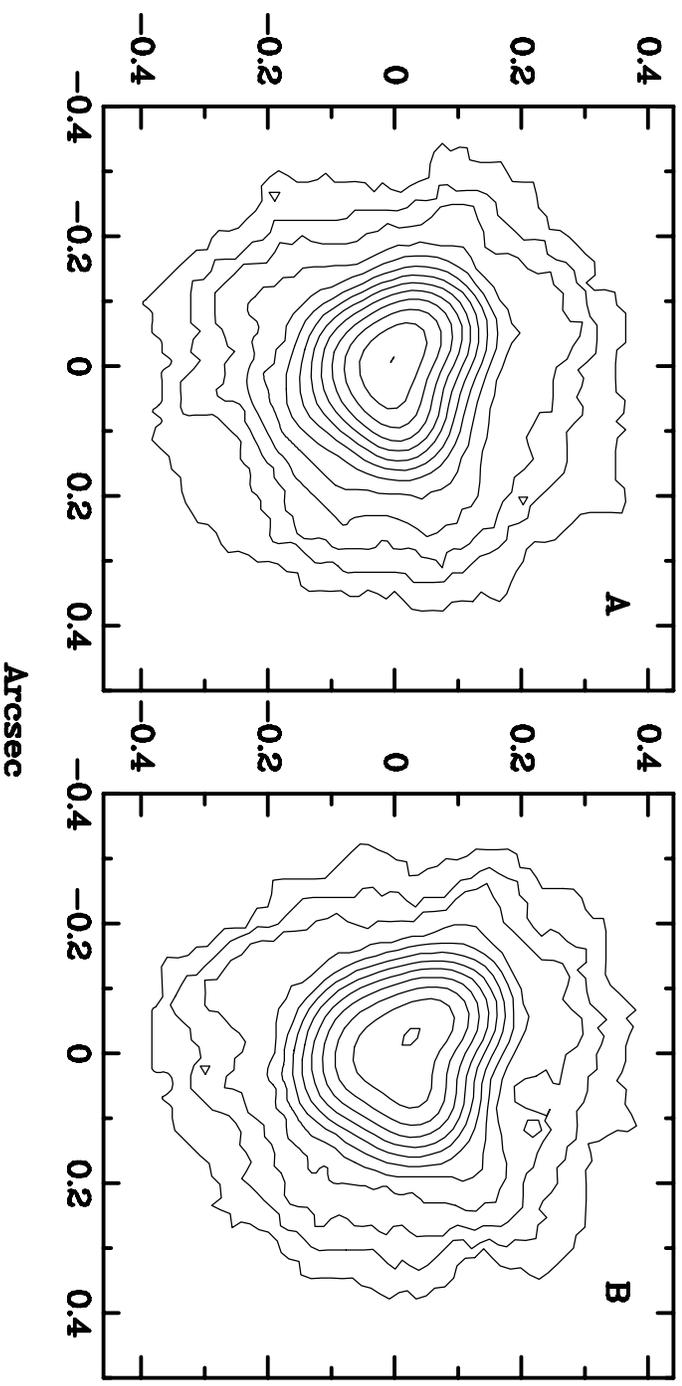
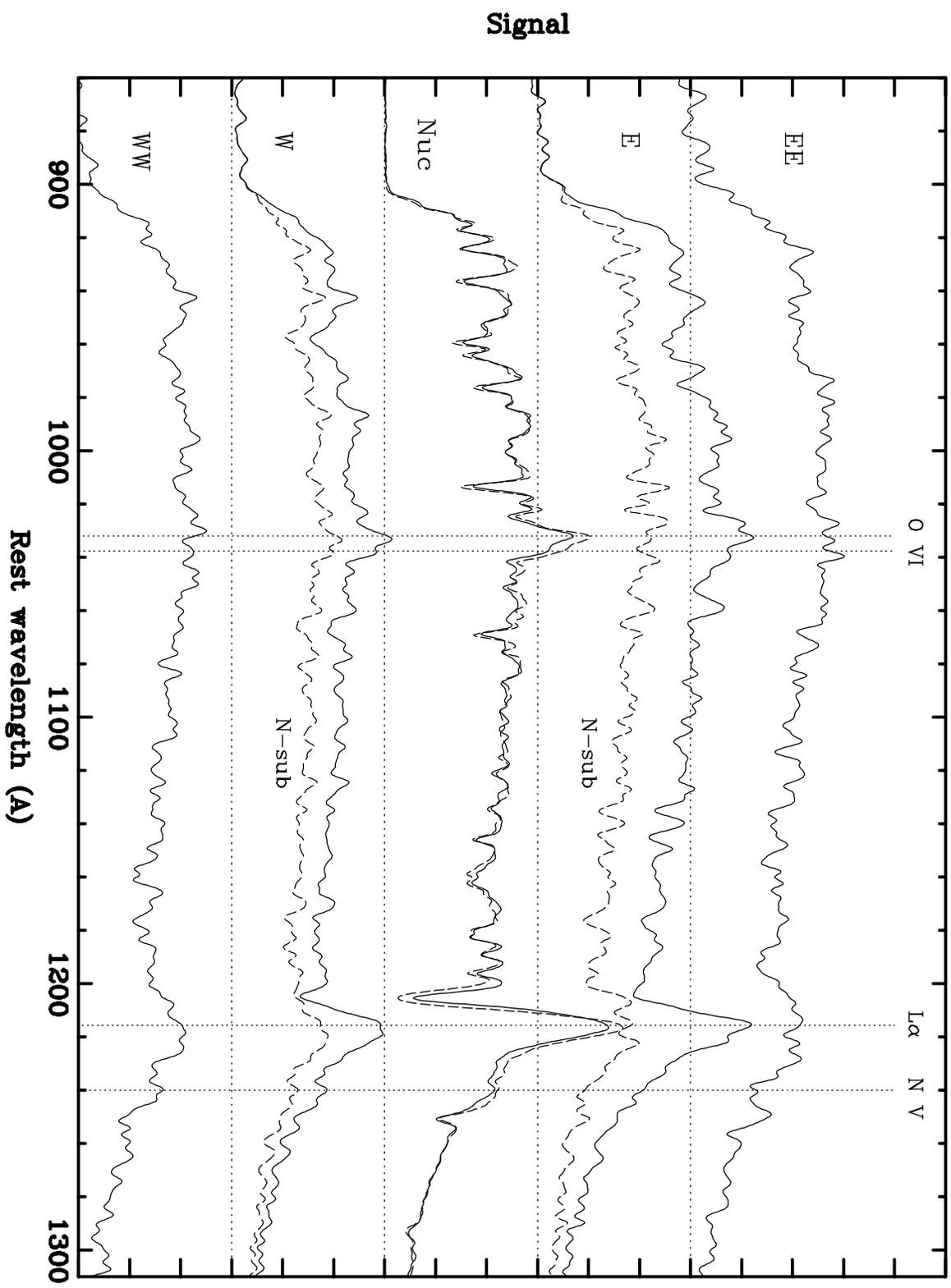
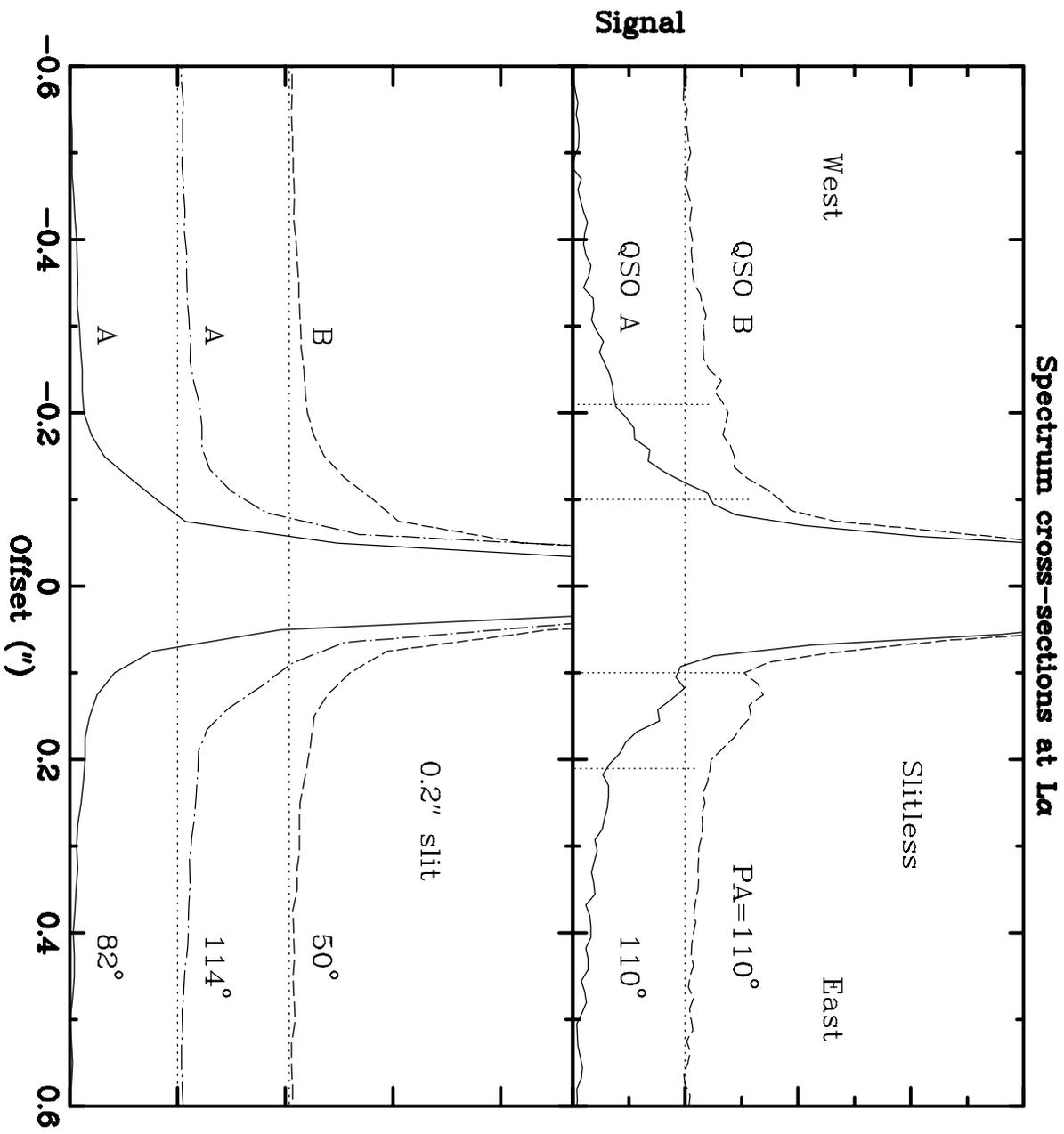
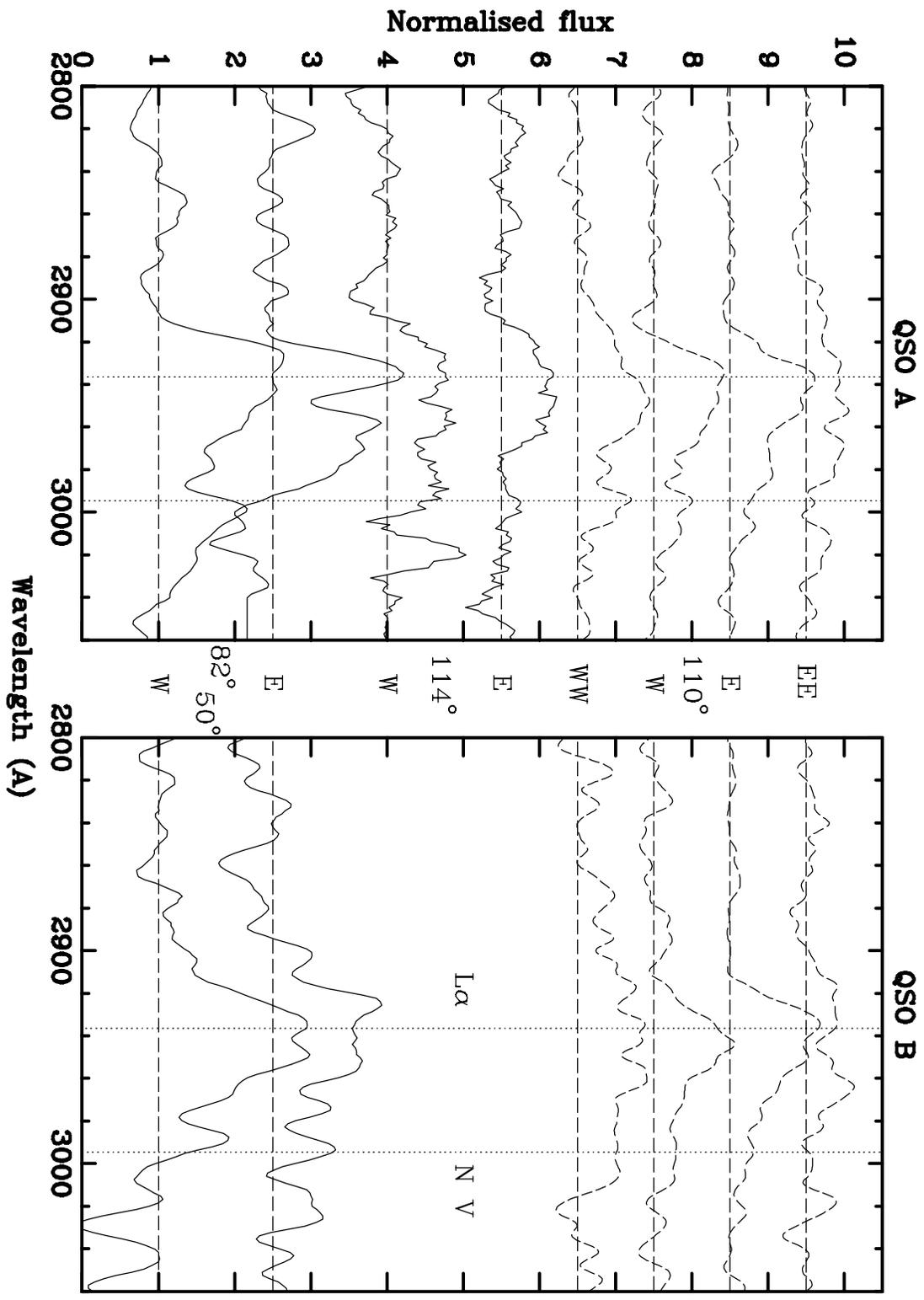
\end{document}